\documentclass[
nofootinbib,
amsmath,amssymb,
prd,
floatfix,
twocolumn
]{revtex4-1}

\usepackage[utf8]{inputenc}
\usepackage{epsf, float, graphicx,dcolumn}
\usepackage{hyperref}
\usepackage{multirow}
\usepackage{booktabs}
\usepackage{siunitx}
\usepackage{soul}
\usepackage{xcolor}
\usepackage{blindtext}
\usepackage{lipsum}

\hypersetup{colorlinks=true,linkcolor=blue,citecolor=blue,filecolor=blue,urlcolor=blue}

\usepackage{scalerel}
\usepackage{tikz}
\usetikzlibrary{svg.path}

\definecolor{orcidlogocol}{HTML}{A6CE39}
\tikzset{
  orcidlogo/.pic={
    \fill[orcidlogocol] svg{M256,128c0,70.7-57.3,128-128,128C57.3,256,0,198.7,0,128C0,57.3,57.3,0,128,0C198.7,0,256,57.3,256,128z};
    \fill[white] svg{M86.3,186.2H70.9V79.1h15.4v48.4V186.2z}
                 svg{M108.9,79.1h41.6c39.6,0,57,28.3,57,53.6c0,27.5-21.5,53.6-56.8,53.6h-41.8V79.1z M124.3,172.4h24.5c34.9,0,42.9-26.5,42.9-39.7c0-21.5-13.7-39.7-43.7-39.7h-23.7V172.4z}
                 svg{M88.7,56.8c0,5.5-4.5,10.1-10.1,10.1c-5.6,0-10.1-4.6-10.1-10.1c0-5.6,4.5-10.1,10.1-10.1C84.2,46.7,88.7,51.3,88.7,56.8z};
  }
}
\newcommand\orcidicon[1]{\href{https://orcid.org/#1}{\mbox{\scalerel*{
\begin{tikzpicture}[yscale=-1,transform shape]
\pic{orcidlogo};
\end{tikzpicture}
}{|}}}}

\begin{document}

\title{The Power of the Cosmic Web}

\author{James Sunseri \orcidicon{0000-0003-4274-2662}}\email{jsunseri@princeton.edu}
\affiliation{Department of Astrophysical Sciences, Princeton University, Princeton, NJ 08540, USA}

\author{Adrian E.~Bayer \orcidicon{0000-0002-3568-3900}}
\affiliation{Department of Astrophysical Sciences, Princeton University, Princeton, NJ 08540, USA}
\affiliation{Center for Computational Astrophysics, Flatiron Institute, 162 5th Avenue, New York NY
10010, USA}

\author{Jia Liu \orcidicon{0000-0001-8219-1995}}
\affiliation{Center for Data-Driven Discovery, Kavli IPMU (WPI), UTIAS, The University of Tokyo, Kashiwa, Chiba 277-8583, Japan}

\date{\today}

\begin{abstract}

We study the cosmological information contained in the cosmic web, categorized as four structure types: nodes, filaments, walls, and voids, using the Quijote simulations and a modified \textsc{nexus+} algorithm. We show that splitting the density field by the four structure types and combining the power spectrum in each provides much tighter constraints on cosmological parameters than using the power spectrum without splitting. We show the rich information contained in the cosmic web structures---related to the Hessian of the density field---for measuring all of the cosmological parameters, and in particular for constraining neutrino mass. We study the constraints as a function of Fourier scale, configuration space smoothing scale, and the underlying field. For the matter field with $k_{\rm max}=0.5\,h/{\rm Mpc}$, we find a factor of $\times20$ tighter constraints on neutrino mass when using smoothing scales larger than 12.5~Mpc/$h$, and $\times80$ tighter when using smoothing scales down to 1.95~Mpc/$h$. However, for the CDM+Baryon field we observe a more modest $\times1.7$ or $\times3.6$ improvement, for large and small smoothing scales respectively.
We release our new \texttt{python} package for identifying cosmic structures \texttt{pycosmmommf} at \url{https://github.com/James11222/pycosmommf} to enable future studies of the cosmological information of the cosmic web. 
\end{abstract}

\maketitle


\section{Introduction}
\label{sec:intro}
With the recent launch of EUCLID\footnote{\url{https://www.euclid-ec.org}} in 2023, and the early release of the DESI One-Percent Survey \citep{DESI_23a, DESI_23b}\footnote{\url{https://www.desi.lbl.gov}}, we are now entering the new era of precision cosmology. Soon, these surveys will be joined by the upcoming Vera Rubin Observatory LSST\footnote{\url{https://rubinobservatory.org}}, PFS\footnote{\url{https://pfs.ipmu.jp}}, Simons Observatory\footnote{\url{https://simonsobservatory.org}}, SPHEREx\footnote{\url{https://spherex.caltech.edu}}, Roman Space Telescope\footnote{\url{https://roman.gsfc.nasa.gov}}, LiteBIRD\footnote{\url{https://www.isas.jaxa.jp/en/missions/spacecraft/future/litebird.html}}, CMB-S4\footnote{\url{https://cmb-s4.org}}, and SKA\footnote{\url{https://www.skatelescope.org}} \citep{Obs_Euclid_22, DESI_2016arXiv161100036D, DESI_II_2019AJ....157..168D, Obs_Rubin_2019, Obs_PFS_14, Obs_Simons_2019, Obs_Spherex_14_1, Obs_Spherex_16_2, Obs_Spherex_18_3, Obs_Roman_19, Obs_LiteBIRD_23, Obs_CMBs4_22, Obs_Cmbs4_22_2, Obs_SKA_23}. 

While standard clustering statistics, such as the power spectrum, remain a reliable choice for understanding the distribution of matter in the Universe and in turn constraining cosmology, there have recently been considerable efforts to extract additional information from clustering by correlating clustering information with information about the local  environment~\citep{cweb_quijote_Bonnaire_22_1, cweb_quijote_Bonnaire_22_2, Paillas_2023_DS, Bayer_2024_Voids}. Alongside this, there has also been work exploring the statistics of extreme environments (clusters and voids) as a means to constrain cosmology~\citep[e.g. the void size function and correlation function][]{voids_massara_15, voids_Kreisch_19, voids_Schuster_19, voids_Contarini_21, cweb_bayer2021detecting, voids_Bayer_23a, voids_DE_Biswas_10, voids_DE_Bonici_23, voids_DE_Davies_21, voids_DE_Euclid_20, voids_DE_Pisani_15, voids_DE_Weinberg_13, Voids_DE_Contarini_22,Thiele2023_BOSSvoids, Golshan:2024lmr}.

Here, we examine the information content of the full cosmic web, which includes not only the most overdense and underdense regions (clusters and voids, respectively), but also the intermediate structures (filaments and walls). Detection of the latter structures has been challenging due to their diffused nature and complex morphology, but has recently been achieved in optical, thermal Sunyaev-Zel'dovich~(tSZ), and X-ray observations of certain fields and specific objects \citep{fils_Bonjean_18, fils_Connor_18, fils_Connor_19, fils_deGraaf_19, fils_Dietrich_05, fils_Eckert_15, fils_Sugawara_17,fils_Planck_13, fils_Tanimura_19, fils_Tanimura_20a, fils_Tanimura_20b, fils_Tittley_01}. The goal of this work is to uncover the potential of measurements like these, to dissect the cosmic web, and in turn constrain cosmological parameters.  

Furthermore, the cosmic web is expected to contain substantial information about the neutrino mass, which is among the top priorities of upcoming surveys \citep{neutrino_Fukuda_98, neutrino_Ahmad_02, neutrino_Araki_05, neutrino_Ahn_06, neutrino_An_12}.  
Massive neutrinos leave fingerprints on large-scale structure observables by altering the growth of structure and the expansion rate~\citep{Review_Workman_22, DETF_albrecht2006report}. 
However, measuring this effect has been exceptionally difficult as neutrinos follow a unique clustering pattern where on scales larger than their free streaming scale they cluster analogously to CDM ($\delta_{\nu} \sim \delta_{\rm cdm}$) and on scales smaller than their free streaming scale they don't cluster ($\delta_{\nu} \sim 0$) due to their thermal velocities preventing gravitational collapse from perturbations. Thus the information regarding neutrino mass lies on small scales where we require models of the non-linear processes in large-scale structure. Recently, this has been made possible with simulations that can accurately model the effects of massive neutrinos on small scales \cite{Bird2018, Villaescusa_Navarro_2018, Arka_2018, Bayer_2021_fastpm, DeRose:2023dmk, sims_Liu2018, sims_Quijote}. 

We explore this potential by dissecting the overdensity field into four distinct structure types (nodes, filaments, walls, and voids) comprising the cosmic web. We identify these structures using a multi-scale morphological filter structure identification scheme implemented in our new \texttt{python} package \texttt{pycosmommf}\footnote{\url{https://github.com/James11222/pycosmommf}} \cite{baryon_Sunseri_23_CosmoMMF} which contains a modified version of the \textsc{nexus+} algorithm \cite{NEXUS_Cautun_2012}. We compute the power spectrum within each of the four components, and then perform a combined analysis on the power spectra.
This technique of splitting the density field into different components and then jointly analyzing the splits is similar in philosophy to the density splits approach \cite{Paillas_2023_DS}, in which the field is split into bins of different density. Instead, here we split based on the physically-intuitive cosmic web components (related to the Hessian of the density field) instead of the density field itself. We apply our code to the Quijote Simulations \cite{sims_Quijote}, a large suite of simulations that vary several cosmological parameters ($\Omega_{\rm m}, \Omega_{\rm b}, h, n_s, \sigma_8$, and  $M_\nu$) relative to a fiducial cosmology as described in Table \ref{tab:sim_table}. 

We extend the analysis of \citep{cweb_quijote_Bonnaire_22_1, cweb_quijote_Bonnaire_22_2}, which used the T-web formalism to define the cosmic web in the total matter field $\delta_m$ with a single effective smoothing scale $R_{\rm eff} = 3.4 \; \mathrm{Mpc}/h$. We employ a different method for detecting cosmic structures which makes use of multiple scales to reflect the high dynamic range of the cosmic web. We investigate the effects of applying different smoothing scales in the process of defining the cosmic web, to understand how much the constraints rely on understanding small-scale physics. Moreover, we examine the information content of the cosmic web in not only the total matter field $\delta_{\rm m} = (1 - f_\nu) \delta_{\rm cb} + f_\nu \delta_{\nu}$, but also the cold dark matter plus baryon field $\delta_{\rm cb}$ -- the latter of which is probed in galaxy surveys and thus gives a closer idea of what will be observed in data from galaxy surveys.\footnote{The neutrino fraction is defined as $f_{\nu} \equiv \Omega_\nu/\Omega_{\rm m}$.} The difference between $\delta_{\rm m}$ and $\delta_{\rm cb}$ is particularly relevant given the results of \cite{Bayer_2022_fake_nus} which showed that while the 3d matter field contains much information about neutrino mass, there is much less information when looking at weak lensing or galaxy surveys.

The paper is structured as follows: we outline the simulations and methodologies we adopt in Section \ref{sec:methods}, discuss our results in Section \ref{sec:results}, and finally conclude in Section \ref{sec:conclusion}.

\section{Methodology}
\label{sec:methods}
In this section, we describe the methodology for our analysis, including the simulations we use~(Section~\ref{subsec:simulations}), our method for tagging cosmic structures~(Section~\ref{subsec: nexus}), and our Fisher analysis~(Section \ref{subsec:fisher}). 

\subsection{Simulations}
\label{subsec:simulations}

We use the Quijote Simulations~\cite{sims_Quijote}, a suite of Cosmological N-Body simulations based on the smoothed particle hydrodynamics code \textsc{gadget-iii}~\cite{sims_Springel2005}. We use a total of 14,000 simulations designed for Fisher analyses, of which, a subset of 8,000 simulations are used for computing covariances and 6,000 are used for derivatives.

The standard fiducial simulations and massless neutrino simulations were initialized with second-order Lagrangian perturbation theory (2LPT). The simulations with massive neutrino were initialized using the Zel’dovich approximation (ZA). A separate set of 500 fiducial simulations initialized with ZA was used for computing derivatives with respect to neutrino mass. 

These simulations have $N_{\rm CDM}$=$512^3$ cold dark matter (CDM) particles, and an additional $N_{\nu}$=$512^3$ neutrino particles for simulations with massive neutrinos, in a $L$=$1 \; \mathrm{Gpc}/ h$ box. Baryonic effects only enter via the linear initial conditions of the simulation -- we do not perform a hydrodynamical simulation. For all simulations, we analyze the $z = 0$ snapshot where the effects of neutrinos are largest. 

The fiducial cosmological parameters are set as \{$\Omega_{\rm m}, \Omega_{\rm b}, h, n_s, \sigma_8, M_\nu, w$\}=\{0.3175, 0.049, 0.6711, 0.9624, 0.834, 0, $-1$\}. The parameters for all the simulations we used are summarized in Table \ref{tab:sim_table}. We do not vary $w$ as there is negligible information regarding $w$ at a single redshift as it only effects the background evolution. To prepare for the cosmic web classification, we create three-dimensional (3D) density grids from individual simulations.\footnote{We use the Piecewise Cubic Spline mass alignment scheme implemented in the \textsc{Pylians3} library \cite{sims_Pylians}.} These regular grids have a resolution of $512^3$ voxels.

\begin{table}
\renewcommand{\arraystretch}{1.5}
\begin{ruledtabular}
\begin{tabular}{cccccccc}
 Name   &$\Omega_m$ & $\Omega_b$ &  $h$   & $n_s$  & $\sigma_8$ & $M_\nu$ & $w$ \\ \hline
Fiducial       &   \underline{0.3175}  &	\underline{0.049}   & \underline{0.6711} & \underline{0.9624} &    \underline{0.834}   &     \underline{0}     &  \underline{-1}  \\
$\Omega_m^{+}$       &  \bf{0.3275}  &	0.049   & 0.6711 & 0.9624 &    0.834   &     0     & $-1$ \\
$\Omega_m^{-}$       &  \bf{0.3075}  &	0.049   & 0.6711 & 0.9624 &    0.834   &     0     & $-1$ \\
$\Omega_b^{++}$      &  0.3175  &	\bf{0.051}   & 0.6711 & 0.9624 &    0.834   &     0     & $-1$ \\
$\Omega_b^{--}$        &  0.3175  &	\bf{0.047}   & 0.6711 & 0.9624 &    0.834   &     0     & $-1$ \\
$h^{+}$   &  0.3175  &	0.049   & \bf{0.6911} & 0.9624 &    0.834   &     0   & $-1$ \\
$h^{-}$   &  0.3175  &	0.049   & \bf{0.6511} & 0.9624 &    0.834   &     0   & $-1$ \\
$n_s^{+}$  &  0.3175  &	0.049   & 0.6711 & \bf{0.9824} &    0.834   &     0   & $-1$ \\
$n_s^{-}$  &  0.3175  &	0.049   & 0.6711 & \bf{0.9424} &    0.834   &     0   & $-1$ \\
$\sigma_8^{+}$  &  0.3175  &	0.049   & 0.6711 & 0.9624 &    \bf{0.849}   &     0   & $-1$ \\
$\sigma_8^{-}$      &  0.3175  &	0.049   & 0.6711 & 0.9624 &    \bf{0.819}   &     0     & $-1$ \\
$M_{\nu}^{+}$  &  0.3175  &	0.049   & 0.6711 & 0.9624 &    0.834   &     \bf{0.1}   & $-1$ \\
\end{tabular}
\caption{\label{tab:sim_table}Cosmological parameters of the 14,000 simulations, a subset of the Quijote simulations, used in our analysis. Neutrino masses are in unit of [eV]. 
}
\end{ruledtabular}
\end{table}

\subsection{Cosmic Structure Identification}
\label{subsec: nexus}

\begin{figure*}[h]
    \centering
    \includegraphics[width=0.8\linewidth]{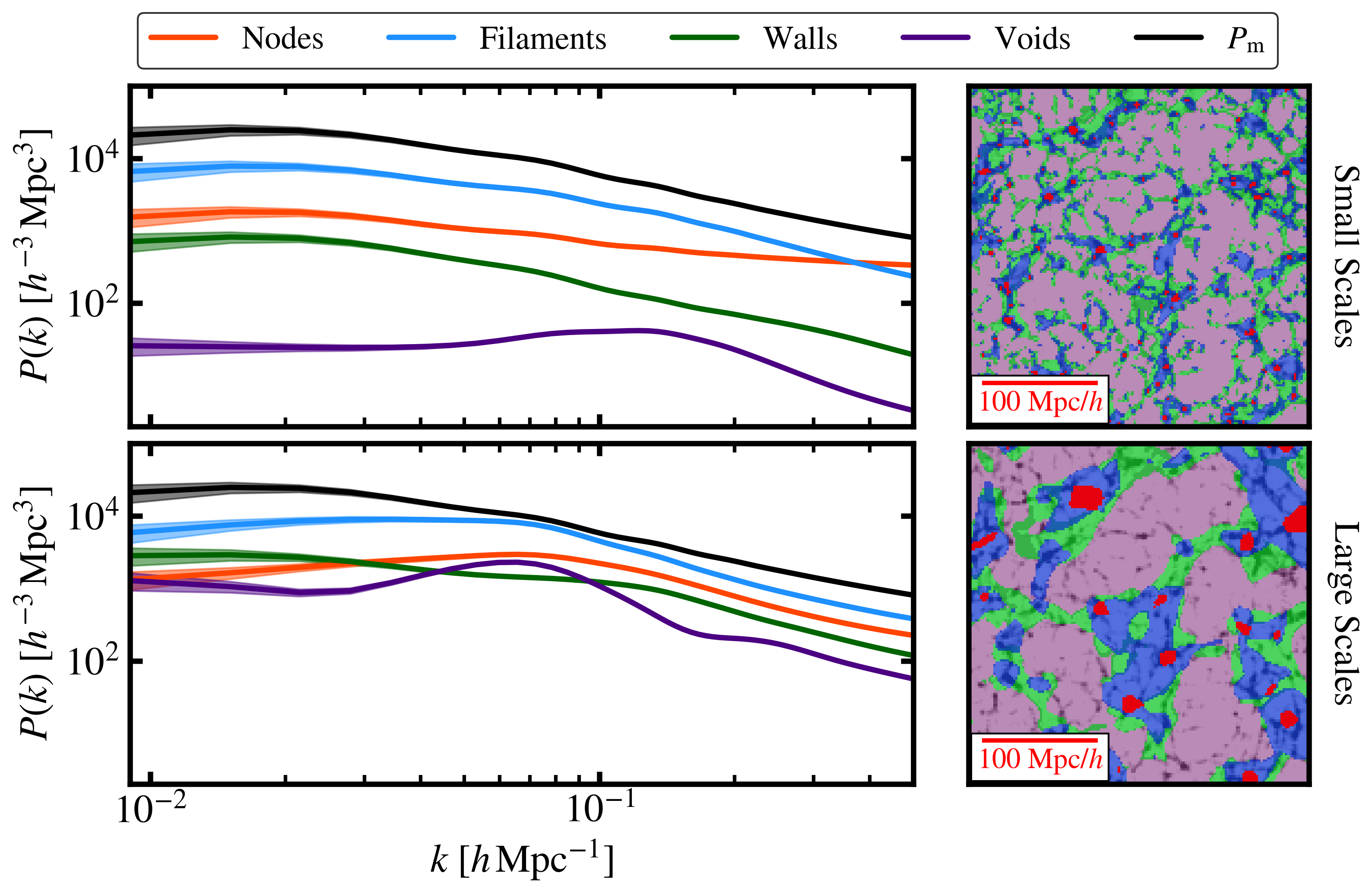}
    \caption{Measured power spectra averaged over all $N_{\rm cov} = 8000$ fiducial simulations with $\pm 1\sigma$ variance. In the left panels we depict the matter power spectrum ($P_{\rm m} = P_{\rm cb}$ for the fiducial simulations) of the full box (black), in nodes (red), filaments (blue), walls (green), and voids (purple). In the top row we used a set of small smoothing scales ($R_{\rm small} < R_{\rm min} = 2 \pi / k_{\rm max}$) and in the bottom row we used a set of large smoothing scales  ($R_{\rm large} \geq R_{\rm min} = 2 \pi / k_{\rm max}$). In the right column we show a subset of the simulated volume with side length $300 \; \mathrm{Mpc}/h$ and thickness of $4 \; \mathrm{Mpc}/h$ containing the tagged structure types nodes (red), filaments (blue), walls (green), and voids (purple) on top of the underlying overdensity field $\delta_{\rm m}$. Limiting the range of scales the structure finding algorithm has access to greatly impacts the morphology of the structures.}
    \label{fig:fiducial_Pk}
\end{figure*}

We use the new python version of our structure identification code \texttt{pycosmommf} \cite{Sunseri_2025_pycosmommf} which was originally written in \texttt{julia} for our previous work \citep{Sunseri_2022_cosmommf, baryon_Sunseri_23a}. It is based on the \textsc{nexus+} multiscale morphological filter algorithm \cite{NEXUS_Cautun_2012}, with a modification to the node detection algorithm. The Quijote simulations do not have a high enough resolution to use the \textsc{nexus+} virialization threshold generating procedure for node detection described in detail in Section 2.1.6 of \cite{NEXUS_Cautun_2012}, thus motivating the modification in this work.
The procedure to identify cosmic structures can be summarized in four main steps:
\begin{enumerate}
    \item \textbf{Smoothing the density field with multiple scales:} In order to accommodate the large dynamic range of structure sizes in the cosmic web \citep{Libeskind_2018_Cosmic_Web}, we examine the density fields in multiple scales. To do so, we compute the smoothed log-density field $f_{R}(\mathbf{x})$ with a Gaussian filter of radius $R$ for a set of smoothing scales $\{R_i\}$.
    We use 2 different sets of smoothing scales in our multiscale morphological filter algorithm to identify morphological structures which we will discuss later in this section.
    \item \textbf{Hessian computation:} To quantify the signature strength of each cosmic structure, for each smoothing scale $R$, we compute the Hessian matrix of $f_{R}(\mathbf{x})$ and its eigenvalues $\lambda_{a}$,  
\begin{align}
    \mathbf{H}_{ij,R}(\mathbf{x}) =& R^2 \frac{\partial^2 f_{R}(\mathbf{x}) }{\partial x_i \partial x_j}\\
    \mathrm{det}(\mathbf{H}_{ij} - \lambda_{a}\mathbf{I}) =& 0, {\rm \; with\;} \lambda_1 \leq \lambda_2 \leq \lambda_3 
\end{align}
    where $a \in \{1,2,3\}$ and $\mathbf{I}$ is the identify matrix. The square of the smoothing scale, $R^2$, is used to normalize the Hessian matrices across a set of different smoothing scales.
    \item \textbf{Structure Classification:} Based on the eigenvalues, we then identify ``node'', ``filament", and ``wall" voxels. Nodes are voxels with $\lambda_1, \lambda_2, \lambda_3<0$, filaments with $\lambda_1, \lambda_2<0$, and walls with $\lambda_1<0$. Therefore, all node voxels also meet the criteria of filament voxels and all filament voxels meet the criteria of walls. For voxels with ambiguous identities, we compare the relative strength of ``filament" versus ``wall" signatures for each of the smoothing scales:
    \begin{align}
    \mathcal{S}^{\rm n}_R(\mathbf{x}) = &  \frac{\lambda_3^2}{|\lambda_1|},\\
    \mathcal{S}^{\rm f}_R(\mathbf{x}) = &  \frac{\lambda_2^2}{|\lambda_1|} \left(1 - \left|\frac{\lambda_3}
{\lambda_1}\right|\right),\\
    \mathcal{S}^{\rm w}_R(\mathbf{x}) =&  |\lambda_1| \left(1 - \left|\frac{\lambda_2}{\lambda_1}\right| \right) \left(1 - \left|\frac{\lambda_3}{\lambda_1}\right| \right).
    \end{align}
    where the final signature strength is taken as the maximum value across all smoothing scales,
\begin{equation}
    \mathcal{S} (\mathbf{x}) = \max\limits_{\textrm{n=1,2,..,}N_{\rm R}} \mathcal{S}_{R_n} (\mathbf{x}).
\end{equation}
We define $N_{\rm R}$ to be the number of smoothing scales in the set. Since the Quijote simulations do not have enough resolution to spatially resolve nodes, we have to manually set a threshold signature value $\mathcal{S}^{\rm n}_{\rm th}$. We choose our threshold value such that our ``nodes" approximate the mass and volume fractions of Friends-of-Friends (FoF) halos \cite{FOF_Press_and_Davis_1982, FOF_Huchra_and_Geller_1982, FOF_Davis_1985} with masses $M \geq 10^{13.3} M_{\odot}/h$ in the fiducial cosmology. The threshold $\mathcal{S}_{\rm th}$ separating filaments and walls is set to be the peak of $\Delta M^2$, which is the mass change with respect to the signature,
\begin{equation}
    \Delta M^2 = \left| \frac{d M^2}{d \log \mathcal{S}} \right|,
\end{equation}
where $M$ is the mass in filaments or walls for a given signature $\mathcal{S}$. After tagging voxels as ``node'' corresponding to $\mathcal{S}^{\rm n} > \mathcal{S}^{\rm n}_{\rm th}$, we then tag any untagged voxels with $\mathcal{S}^{\rm f}>\mathcal{S}^{\rm f}_{\rm th}$ as ``filament". Lastly, the remaining (non-node and non-filament) voxels with $\mathcal{S}^{\rm w}>\mathcal{S}^{\rm w}_{\rm th}$ are tagged as ``wall".\footnote{Detailed discussions on optimal filament and wall separation can be found in Appendix A of the \textsc{nexus+} paper~\cite{NEXUS_Cautun_2012}.} Any voxels left unidentified are tagged as ``void". The end result can be seen in the right column of Figure \ref{fig:fiducial_Pk}.
\end{enumerate}

We use 2 sets of smoothing scales in our analysis: $R_{\rm small} = R_0 \times \{ 1, \sqrt{2}, 2, ..., 4\sqrt{2}\}$ where $R_0 = (1 \; \mathrm{Gpc}/h ) / 512 \approx 1.95 \; \mathrm{Mpc}/h$ is the Quijote particle cell spacing (we refer to this as ``small scales'') and $R_{\rm large} = R_{\rm min} \times \{ 1, \sqrt{2}, 2\}$, where $R_{\rm min} = 2 \pi /0.5 \approx 12.5 \; \mathrm{Mpc}/h$ (we refer to this as ``large scales''). The choice in smoothing scales has a large impact on the resulting structures of the cosmic web as can be seen in Figure \ref{fig:fiducial_Pk} and their summary statistics\footnote{Using different smoothing scales requires that we use different threshold values for identifying nodes in the cosmic web. For small smoothing scales we use $\mathcal{S}_{\rm th}^{\rm c} = 0.01$ and for large smoothing scales we use $\mathcal{S}_{\rm th}^{\rm c} = 0.1$. Both structure identification schemes produce similar mass and volume fractions.}. 
We choose these two sets of scales, because while many different cosmic web classifiers commonly use smoothing scales $R \lesssim 10 \; \mathrm{Mpc}/h < R_{\rm min}$ to identify structures in the cosmic web \citep{Libeskind_2018_Cosmic_Web, Ramsoy_2021_CWEB, Espinosa_2023_CWEB, Espinosa_2024_CWEB, Wang_2024_CWEB}, the power spectra of the Quijote simulations used in this work have been shown to be converged only up to $k_{\rm max} = 0.5 \; h \rm Mpc^{-1}$ \cite{sims_Quijote}, which roughly corresponds to a minimum physical scale of $R_{\rm min} = 2 \pi / k_{\rm max} \approx 12.5 \; \mathrm{Mpc}/h$. While the configuration space clustering of particles may be converged all the way down to the initial inter-particle spacing $R_0$, it is instructive to investigate the implications of using larger smoothing scales as it relates to the information content of the Hessian of the density field.\footnote{We note that when we use large smoothing scales, we are no longer describing the ``cosmic web'' as it is commonly described in the literature \cite{Libeskind_2018_Cosmic_Web}, but instead morphological structures derived from Hessian matrix of the density field with similar characteristics to the conventional cosmic web.} 
    
\begin{figure*}
    \centering
    \includegraphics[width=0.8\linewidth]{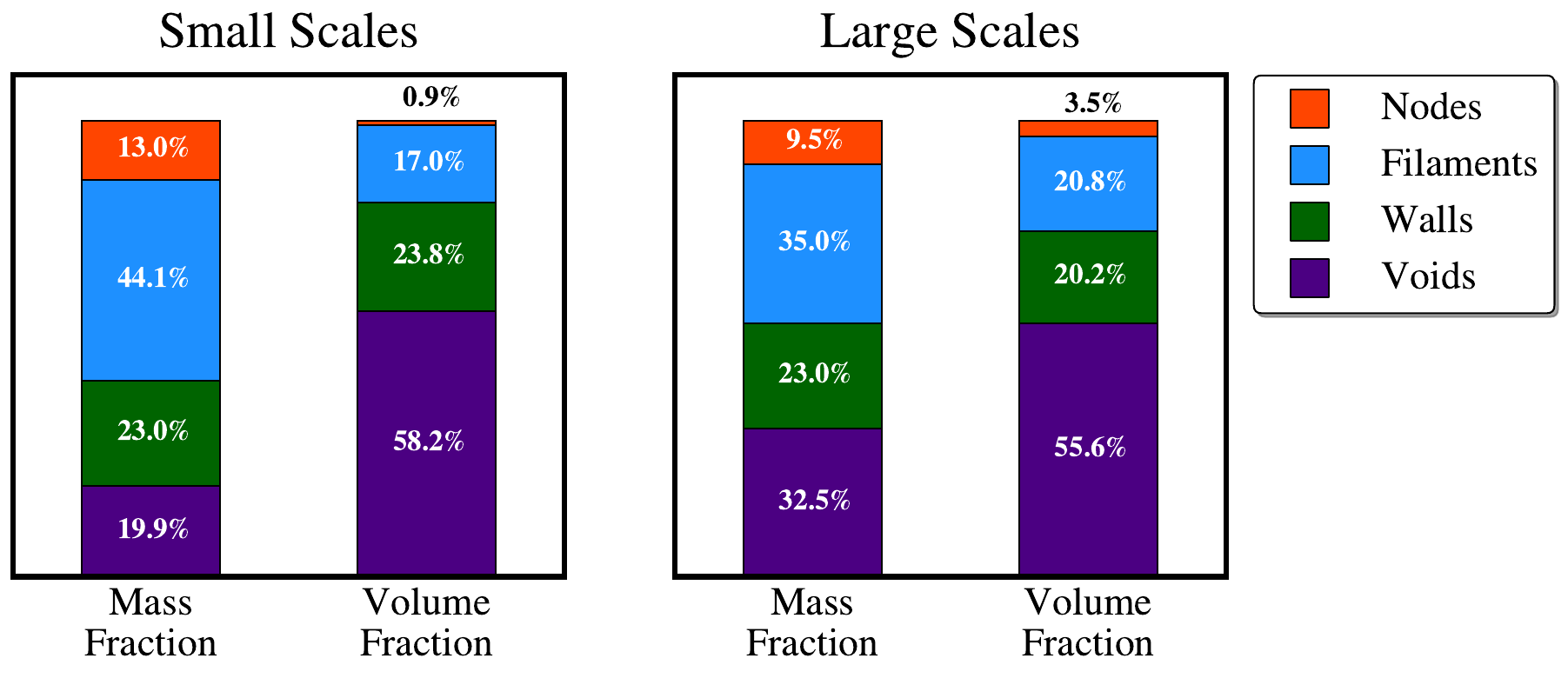}
    \caption{We show the mass and volume fractions of our cosmic web structures (nodes, filaments, walls, and voids) between our 2 sets of smoothing scales. There is variance in mass and volume fractions between structure finding algorithms \cite{Libeskind_2018_Cosmic_Web}, but both of our choices of smoothing scales yield mass and volume fractions within the expected scatter from different methods.}
    \label{fig:fractions}
\end{figure*}

As a point of comparison and consistency between our 2 choices of smoothing scales in defining the cosmic web, we compare mass and volume fractions of the 4 structure types. In Figure \ref{fig:fractions}, we show the measured mass and volume fractions over 8000 fiducial simulations and see that the small and large sets of smoothing scales do yield consistent fractions within the expected variance using the \textsc{nexus+} method  \citep{Cautun_2014_evolution_of_cweb, Hellwing_2021_cweb} and various structure finding algorithms (see \cite{Libeskind_2018_Cosmic_Web} for a review of these methods). 

We note the largest difference between the two cosmic web definitions is the mass found in filaments. When using small smoothing scales we see 10\% more mass in filaments than when using large smoothing scales, this is to be expected and can be understood when looking at the right side of Figure \ref{fig:fiducial_Pk}. We can see that in this case many overdense regions which were identified as filaments with access to small smoothing scales are tagged as voids because the structure finder is limited to scales larger than $2\pi/k_{\rm max}$. 

\subsection{Fisher Information}
\label{subsec:fisher}
To calculate the embedded information within the cosmic web we use the Fisher matrix formalism \citep{Tegmark_1997_Fisher, Heavens_2009_Fisher, Verde_2010_Fisher_Stats}. We define the Fisher matrix as 
\begin{equation}
    F_{ij} = - \left< \frac{\partial^2 \mathcal{L}}{\partial \theta_i \partial \theta_j} \right> \; ,
\end{equation}
where $\mathcal{L}$ is the likelihood and $\vec{\theta}$ corresponds to the parameter vector of the model \cite{Fisher_1925}. If we assume that the region around the maximum likelihood is well approximated by a multivariate Gaussian distribution, then one can write down an analytical form for the Fisher matrix
\begin{multline}
    F_{ij} = \frac{1}{2} \left[\frac{\partial \vec{O}}{\partial \theta_i} C^{-1} \frac{\partial \vec{O}^T}{\partial \theta_j} + \frac{\partial \vec{O}}{\partial \theta_j} C^{-1} \frac{\partial \vec{O}^T}{\partial \theta_i}\right] \\ + \frac{1}{2} \mathrm{Tr}\left[C^{-1} \frac{\partial C}{\partial \theta_i} C^{-1} \frac{\partial C}{\partial \theta_j} \right] \; ,
\label{eq:Fisher}
\end{multline}
where $\vec{O}$ corresponds to the observable vector and $C$ is the covariance matrix. Since we are assuming a Gaussian likelihood, we ignore the covariance derivative term in Eq. \ref{eq:Fisher} to prevent underestimation of errors \cite{Carron_2013_Fisher_Cov}. Lastly, with the Fisher matrix we can estimate the lower bound error of our cosmological parameters as 
\begin{equation}
    \sigma(\theta_i) \equiv \sqrt{\left(F^{-1}\right)_{ii}} \; .
\end{equation}
\subsubsection{Observable Vector}

Upon creating structure filters for nodes, filaments, walls, and voids as described in Section \ref{subsec: nexus}, we then isolate structures in the density field that correspond to a specific structure type by setting the density $\rho = 0$ in all other voxels in the periodic box. We then compute the (auto) power spectrum using the \textsc{Pylians3} library \cite{sims_Pylians}
\begin{equation}
    \langle \delta^*(\mathbf{k}) \delta(\mathbf{k'})\rangle = (2 \pi)^3 P_{\delta}(k) \delta^{(D)} (\mathbf{k} - \mathbf{k'}) \; ,
\end{equation}
where $\delta$ is the overdensity field and $\delta^{(D)}$ is the Dirac delta function. We denote the power spectrum of the matter density field as $P_{\rm m}(k)$ and the power spectrum of the CDM+Baryon overdensity field as $P_{\rm cb}(k)$. Our observable vector can be written as 
\begin{multline}
    \vec{O} = \{P_{\delta}^{\rm n}(k_{1}), ..., P_{\delta}^{\rm n}(k_{n}),  P_{\delta}^{\rm f}(k_{1}), ..., P_{\delta}^{\rm f}(k_{n}), \\
    P_{\delta}^{\rm w}(k_{1}), ..., P_{\delta}^{\rm w}(k_{n}), P_{\delta}^{\rm v}(k_{1}), ..., P_{\delta}^{\rm v}(k_{n})\} \; , 
\end{multline}
and the parameter vector as
\begin{equation}
    \vec{\theta} = \{ \Omega_{\rm m}, \Omega_{\rm b}, h, n_{\rm s}, \sigma_8, M_{\nu} \} \; .
\end{equation}
We do not study the constraining power on the dark energy equation of state parameter, as there is no constraining power when considering a fixed redshift snapshot, as we do here. 
We refer to $P_{\delta}^\alpha(k)$ as the (auto) power spectrum of an overdensity field $\delta$ and structure type $\alpha = \{\rm n, f, w, v\}$ (nodes, filaments, walls, and voids). The total length of the observable vector $\vec{O}$ is $4N$ where $N$ is the number of bins in the power spectrum measurement. We bin our power spectra measurements into $N=79$ log-spaced wavenumber bins with the largest wavenumber being $k^{\rm fid}_{\rm max} = 0.5 \; h \rm Mpc^{-1}$. 

\subsubsection{Derivatives \& Covariance}
We compute the derivatives in the Fisher matrix using the central difference scheme 
\begin{equation}
    \frac{\partial \vec{O}}{\partial \theta_i} = \frac{\vec{O}(\theta_i + \Delta \theta_i) - \vec{O}(\theta_i - \Delta \theta_i)}{2 \Delta \theta_i} \; ,
\label{eq:central_diff}
\end{equation}
for all cosmological parameters $\{\Omega_{\rm m}, \Omega_{\rm b}, h, n_{\rm s}, \sigma_8\}$ except the total neutrino mass. For the standard $\Lambda$CDM parameters we averaged over $N_{\rm der} = 500$ simulations for each term Eq. \ref{eq:central_diff} (totaling 1000 simulations per derivative). The central difference scheme can not be used for neutrino mass because it would require simulations with negative neutrino mass. For this reason we need to use a forward difference scheme. In our analysis, we found the most stable forward difference scheme to be a 2-point forward difference scheme 
\begin{equation}
    \frac{\partial \vec{O}}{\partial M_\nu} = \frac{\vec{O}(\Delta M_\nu) - \vec{O}(M_\nu=0)}{\Delta M_\nu} \; ,
\end{equation}
where $\Delta M_\nu = 0.1$\footnote{We explored using the $M_{\nu}^{++}$ and $M_{\nu}^{+++}$ simulations with 3 and 4-point forward difference schemes and found them to be less stable in this Fisher analysis.}. This scheme makes use of the fiducial simulation $M_\nu = 0$ initialized with ZA initial conditions to match with the $M_\nu = 0.1$ simulations.

We use $N_{\rm cov} = 8000$ fiducial cosmology simulations to estimate the covariance matrix. The covariance matrix can be estimated as
\begin{equation}
    C_{i j} = \langle (O_{i} - \langle O_{i} \rangle) (O_{j} - \langle O_{j} \rangle) \rangle \; ,
\end{equation}
where $\langle \rangle$ denotes an average over the 8000 fiducial simulations. All results presented in this work make use of the full $N_{\rm cov} = 8000$ fiducial simulations for covariance estimation and $N_{\rm der} = 500$ derivative simulations to evaluate each term in the derivative with respect to cosmological parameter.

\section{Results}
\label{sec:results}
\begin{table*}
\renewcommand{\arraystretch}{1.5}
\begin{ruledtabular}
\begin{tabular}{l|cccccc}
\toprule
 Statistic on $\delta_{\rm m}$ [Publication] & $\sigma(\Omega_{m})$ & $\sigma(\Omega_{b})$ & $\sigma(h)$ & $\sigma(n_{s})$ & $\sigma(\sigma_{8})$ & $\sigma(M_{\nu})$ [eV] \\ \hline \hline
\midrule
$M(k)$ - \citep[Massara et.\textit{al} 2021,][]{Massara_2021_Marked} & 0.013 & 0.010 & 0.095 & 0.044 & \textbf{0.002} & 0.016 \\
WST - \citep[Valogiannis \& Dvorkin 2022,][]{Valogiannis_2023_WST} & 0.014 & 0.012 & 0.103 & \textbf{0.031} & \textbf{0.001} & \textbf{0.008} \\
$P(k)$ + HMF + VSF - \citep[Bayer et.\textit{al} 2021,][]{cweb_bayer2021detecting} & \textbf{0.006} & 0.037 & 0.230 & 0.100 & 0.007 & 0.096 \\
(T-Web) $P_{\rm m}^{\rm comb.}$ - \citep[Bonnaire et.\textit{al} 2021,][]{cweb_quijote_Bonnaire_22_1} & 0.012 & \textbf{0.009} & \textbf{0.079} & 0.032 & 0.005 & 0.058 \\ \hline \hline 
$P_{\rm m}$ & 0.098 & 0.039 & 0.508 & 0.483 & 0.013 & 0.826 \\
$P_{\rm m}^{\rm comb.}$ (Small Scales) [this work] & \textbf{0.011} & \textbf{0.008} & \textbf{0.069} & \textbf{0.026} & \textbf{0.002} & \textbf{0.010} \\
$P_{\rm m}^{\rm comb.}$ (Large Scales) [this work] & 0.017 & 0.012 & 0.109 & 0.050 & \textbf{0.002} & 0.040 \\
\bottomrule
\end{tabular}
\end{ruledtabular}
\caption{Marginalized $1\sigma$ error compared to other statistics found in the literature for the $\delta_{\rm m}$ field. We compare our results to the Marked power spectrum $M(k)$, Wavelet Scattering Transform (WST), Halo Mass Function + Void Size Function (HMF+VSF), and the combined power spectrum from correlating structures in the cosmic web (identified with the T-Web algorithm, marginalized over nuisance parameters). We show the matter power spectra constraints $P_{\rm m}$ as a reference. Our most competitive constraints, coming from using smaller smoothing scales than $k_{\rm max}$ on the matter field, show comparable constraints to other state-of-the-art statistics found in the literature. We bold the top 2 statistics in each column, showing that this work provides top-2 results for all parameters, while other statistics are optimal only for specific parameters.}
\label{tab:constraints_m}
\end{table*}

\begin{table*}
\renewcommand{\arraystretch}{1.5}
\begin{ruledtabular}
\begin{tabular}{l|cccccc}
\toprule
 Statistic on $\delta_{\rm cb}$ [Publication] & $\sigma(\Omega_{m})$ & $\sigma(\Omega_{b})$ & $\sigma(h)$ & $\sigma(n_{s})$ & $\sigma(\sigma_{8})$ & $\sigma(M_{\nu})$ [eV] \\ \hline \hline
\midrule
$M(k)$ - \citep[Massara et.\textit{al} 2021,][]{Massara_2021_Marked} & \textbf{0.016} & \textbf{0.009} & \textbf{0.083} & \textbf{0.035} & \textbf{0.026} & \textbf{0.450} \\
WST - \citep[Valogiannis \& Dvorkin 2022,][]{Valogiannis_2023_WST} & \textbf{0.016} & 0.012 & 0.103 & \textbf{0.029} & \textbf{0.017} & \textbf{0.29} \\ \hline \hline 
$P_{\rm cb}$ & 0.066 & 0.018 & 0.198 & 0.133 & 0.104 & 1.757 \\ 
$P_{\rm cb}^{\rm comb.}$ (Small Scales) [this work] & 0.023 & \textbf{0.009} & \textbf{0.086} & 0.045 & 0.028 & 0.488 \\
$P_{\rm cb}^{\rm comb.}$ (Large Scales) [this work] & 0.039 & 0.013 & 0.123 & 0.064 & 0.062 & 1.042 \\
\bottomrule
\end{tabular}
\end{ruledtabular}
\caption{Marginal $1\sigma$ error compared to other statistics found in the literature for the $\delta_{\rm cb}$ field. We compare our results to the Marked power spectrum $M(k)$ and the Wavelet Scattering Transform (WST). We show the CDM+Baryon power spectra constraints as a reference in the middle of the table. Our most competitive constraints, coming from using smaller smoothing scales than $k_{\rm max}$, show comparable constraints to other state-of-the-art statistics found in the literature. We bold the top 2 statistics in each column, showing that this work provides top-2 results for $\Omega_{\rm b}$ and $h$ but not for the other cosmological parameters.}
\label{tab:constraints_cdm}
\end{table*}

We use the Fisher formalism to explore the information content within the cosmic web. We perform four analyses by modifying three key considerations:
\begin{itemize}
    \item \textbf{Smoothing scales:} we identify structures using smoothing scales $R_{\rm small} <2 \pi / k_{\rm max}^{\rm fid} \approx 12.5 \; \mathrm{Mpc} / h $ and structures identified using smoothing scales $R_{\rm large} \geq 2 \pi/k_{\rm max}^{\rm fid}$, where $k_{\rm max}^{\rm fid} = 0.5 \; h/{\rm Mpc}$ is the fiducial $k$ cut we employ for the power spectrum. Explicitly, our set of small smoothing scales is $R_{\rm small} = \{2, 2\sqrt{2}, 4, ..., 8\sqrt{2}\} \; h^{-1} \rm Mpc$ and our set of large smoothing scales is $R_{\rm large} = \{12.5, 12.5 \sqrt{2}, 25 \} \; h^{-1} \rm Mpc$.
    \item \textbf{Matter versus CDM+Baryons:} we measure clustering of the cosmic web in the $\delta_{\rm m}$ and $\delta_{\rm cb}$ field.
    \item \textbf{Fourier scales:} we investigate the impact of varying the maximum Fourier scale $k$ of the power spectra.
\end{itemize}

By exploring each of these different scenarios, we can unpack the information content within the cosmic web, and understand what can realistically be observed by upcoming surveys. All of our marginalized constraints are listed in Table \ref{tab:constraints_m} and Table \ref{tab:constraints_cdm}.

We focus on the $M_{\nu}$--$\sigma_8$ plane as breaking this degeneracy has been shown to be key in obtaining optimal constraints \cite{voids_Bayer_23a}. We include results for other cosmological parameters in Appendix \ref{subsec:A2}.

\begin{figure*}
    \centering
    \includegraphics[width=0.8\linewidth]{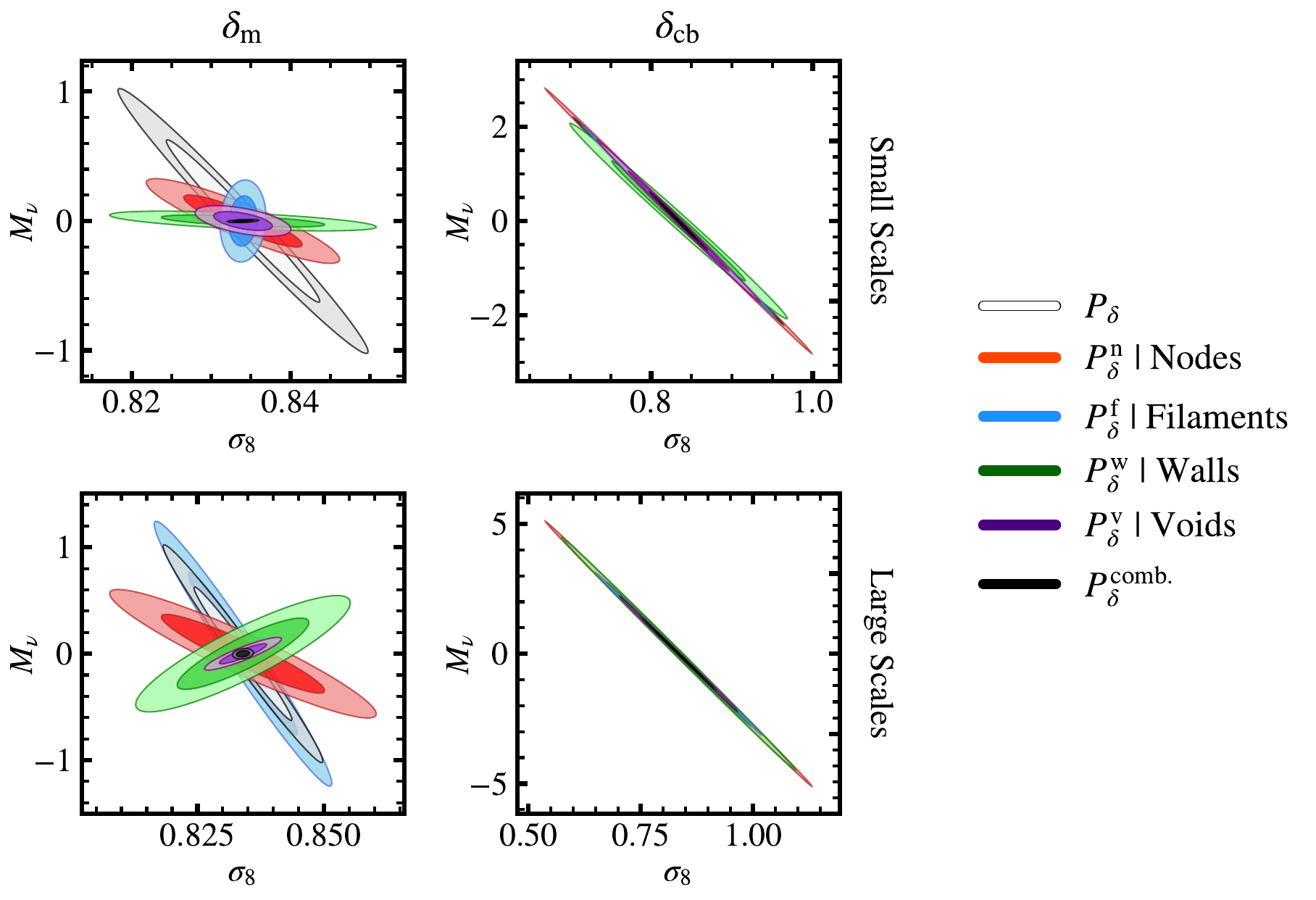}
    \caption{The $M_{\nu}$-$\sigma_8$ plane of the confidence contours for our cosmological parameters across all 4 different analyses. We highlight this plane as it best demonstrates the impact of choices that can be made when using clustering of the cosmic web as a summary statistic. All panels show the $68\%$ (darker shades) and $95\%$ (lighter shades) confidence contours from the standard full power spectrum (white), each structure type --- nodes (red), filaments (blue), walls (green), and voids (purple) --- and the cross correlation of all cosmic structures (black). The left column shows the results of analyses which use the $\delta_{\rm m}$ field and the right column shows contours from analyses using the $\delta_{\rm cb}$ field which can be probed by galaxy surveys. The top row shows analyses which used smoothing scales smaller than $k_{\rm max}$ and the bottom row contains analyses which used smoothing scales larger than $k_{\rm max}$. 
    }
    \label{fig:2x2}
\end{figure*}

Firstly, we consider the case of small smoothing scales on the $\delta_{\rm m}$ field, depicted in the upper left panel of Figure \ref{fig:2x2}. In this case, constraints are significantly tighter than the corresponding power spectrum of the entire field. We see each structure has complimentary degeneracies coming from their individual power spectra $P^{\alpha}_{\rm m}$ that when combined leads to a factor of $\sim 80$ improvement over the standard matter power spectrum on constraining the neutrino mass. The most complimentary structures appear to be filaments and walls. In this analysis scenario, we also see an apparent aligned degeneracy between voids and nodes although both structures have tighter constraints than the power spectrum of the entire field. We expect that if we had a larger dynamic range of scales, we would see the nodes and voids contours rotate in this plane as more large scales are added into the structure finder (this can be seen in the bottom left panel of Figure \ref{fig:2x2}). 

Secondly, we consider the case of large smoothing scales on the $\delta_{\rm m}$ field, depicted in the lower left panel of Figure \ref{fig:2x2}. We can see that constraints are four times worse for the combined analysis of all structures compared to the small smoothing scale case, but still resulting in a factor of  $\sim 20$ improvement on constraining the neutrino mass compared to the standard matter power spectrum. Most notable differences between the small scale and large scale analyses on the matter field is the size and orientation of the constituent structure contours. All contours except for voids are inflated in size, with the most extreme change being for filaments. We see that walls and voids have degeneracies that are aligned while nodes and filaments have dissimilar degeneracies. An interesting difference in the large smoothing scale analysis compared to the small smoothing scale analysis is that the filament contour is comparable in both size and orientation to the matter power spectrum contour. This can be understood when looking closely at the right side of Figure \ref{fig:fiducial_Pk}. We can see that with larger smoothing scales the structure finder does not clearly trace the density field resulting in large ``puffy'' definitions of nodes and filaments compared to the small smoothing scales case. In that case structures are more tightly wrapped around overdense regions. With such loose definitions of filaments, the range of overdensities within filaments is much larger and thus more closely resembles the matter power spectrum contour\footnote{We note that the mass and volume fractions of both definitions of cosmic structures are comparable and serve as a baseline of commonality.}.

Thirdly, we consider the case of small smoothing scales on the $\delta_{\rm cb}$ field, depicted in the upper right panel of Figure \ref{fig:2x2}. We can see a striking difference from the matter field counterpart. We note that all the contours are stretched out and aligned in the $M_{\nu}-\sigma_8$ plane. We do notice that the walls contour has a slight off-axis tilt compared to all the other structures which could be due to lack of exact convergence or a physical effect -- this could be better studied with higher resolution simulations. There is no longer a strong degeneracy breaking between the different cosmic web components, and the combined analysis of all structure types in this analysis yields a marginalized constraint on neutrino mass that is $\sim 3.6$ times better than the CDM+Baryon power spectrum.

Finally, we consider large smoothing scales on the $\delta_{\rm cb}$ field, seen in the bottom right panel of Figure \ref{fig:2x2}. This provides the least constraining power; by eliminating the small smoothing scales, we effectively reduce the amount of non-linear information the cosmic web statistic has access to, and thus our uncertainties inflate. In this scenario, the contours of all structures are aligned and stretched out even further than the third scenario. The marginalized error on $M_\nu$ from cross correlating all structures in this scenario is $\sim 1.7$ times better than the standard CDM+Baryon power spectrum.

\begin{figure}[h]
    \centering
    \includegraphics[width=\linewidth]{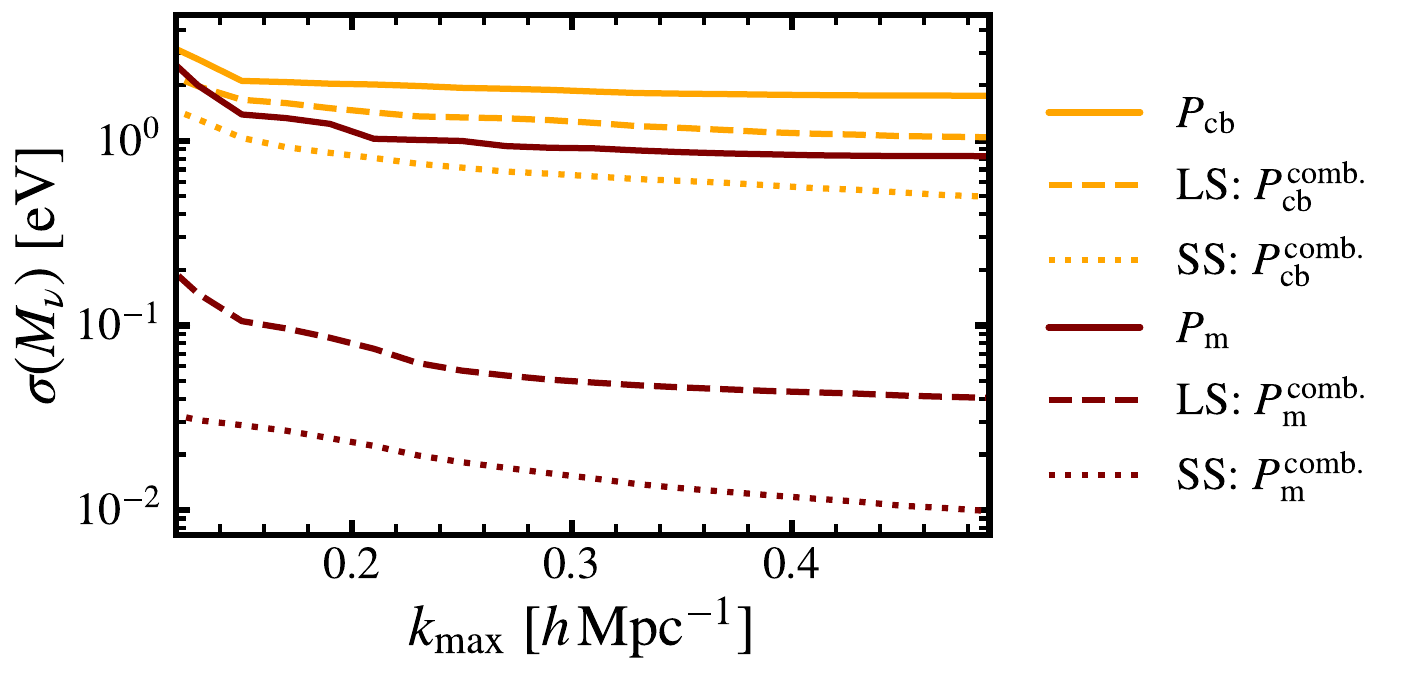}
    \caption{Marginalized $1 \sigma$ error of $M_\nu$ as a function of $k_{\rm max}$. We show the standard power spectrum of the matter field in maroon and the $\delta_{\rm cb}$ field (probed with galaxy surveys) in orange. We compare the marginalized errors from the standard power spectrum to the cosmic web power spectrum (correlating the power spectra of all structures). We have 4 different analysis curves: large smoothing scales with the $\delta_{\rm cb}$ field (orange, dashed), small smoothing scales with the $\delta_{\rm cb}$ field (orange, dotted), large smoothing scales with the matter field (maroon, dashed), and small smoothing scales with the matter field (maroon, dotted).}
    \label{fig:kmax}
\end{figure}

We now compare our results to other literature which used the Quijote simulations to constrain cosmological parameters in Tables \ref{tab:constraints_m} and \ref{tab:constraints_cdm} for the matter and CDM+baryon fields respectively. We find that our marginalized constraints when we use small smoothing scales are competitive with the marked power spectrum \cite{Massara_2021_Marked}, Wavelet Scattering Transform (WST) \cite{Valogiannis_2023_WST}, Halo Mass Function (HMF) + Void Size Function (VSF) \cite{voids_Bayer_23a}, and a similar T-web analysis \cite{cweb_quijote_Bonnaire_22_1} for both the matter and CDM+baryon fields (where available) for all parameters. Our marginalized constraints on $\Omega_{\rm b}$, $h$, and $n_{s}$ are the tightest of all probes considered in the case of small smoothing scales on $\delta_{\rm m}$. The marked power spectrum and WST fold in very small scale information from the density field during their computation, so comparing our small scale analysis is the most instructive comparison to make. When we limit the structure finder to use large smoothing scales our constraints reduce, which would likely also be the case if the smoothing scales were increased for the other statistics---it would be interesting to explore a scale to scale comparison of all the statistics, but this is beyond the scope of this work.

We further investigate the impact of Fourier scale cut when computing the power spectra on our marginalized error of neutrino mass by comparing our different analysis scenarios as a function of $k_{\rm max}$ (see Figure \ref{fig:kmax}). Constraining power saturates for the standard power spectrum of the $\delta_{\rm cb}$ field around $k_{\rm max} = 0.2 \; h/\rm Mpc$ while the matter power spectrum saturates around $k_{\rm max} = 0.3 \; h/\rm Mpc$. In all cases the combined analysis of the cosmic web structure power spectra outperforms the standard power spectrum. 

Through this analysis, we show the most dramatic reduction in information content of the cosmic web comes from considering $\delta_{\rm cb}$ instead of $\delta_{\rm m}$, rather than the scales used to define the cosmic web ($R_{\rm small}$ vs. $R_{\rm large}$). Between the standard $P_{\rm m}$ and $P_{\rm cb}$, we find a factor of $\sim 2$ reduction in constraining power on neutrino mass which is in line with what was shown in \cite{Bayer_2022_fake_nus}. This effect is exacerbated in our combined cosmic web clustering analyses where we find a factor of $\sim 25-50$ reduction in constraining power going from $\delta_{\rm m}$ to $\delta_{\rm cb}$ (at least an order of magnitude larger). The significance of this result is that for galaxy surveys, which trace the underlying cold dark matter field, the improvement in constraining power on neutrino mass is more modest than the total matter field would imply. This is in line with \cite{Bayer_2022_fake_nus} which showed this difference in information content between $\delta_{\rm m}$ and $\delta_{\rm cb}$ applies at the field level, and thus to any summary statistic.

\section{Conclusion}
\label{sec:conclusion}
In this work, we use a modified version of the \textsc{nexus+} algorithm to identify cosmic web structures---nodes, filaments, walls, and voids---in the Quijote simulations (e.g. Figure \ref{fig:fiducial_Pk}). We investigate the information content of these constituent comic structures by jointly analyzing the power spectrum in each environment. We build on the results of \cite{cweb_quijote_Bonnaire_22_1} by using \textsc{nexus+} instead of T-Web for cosmic web definition, and crucially, by investigating the impact on the constraining power of the underlying field ($\delta_{\rm m}$ vs. $\delta_{\rm cb}$), the smoothing scales used in cosmic web algorithm, and the maximum Fourier scale of the power spectrum. 
\newline \newline
Our key findings are listed below:
\begin{itemize}
    \item Jointly analyzing the matter power spectra of different cosmic structures (nodes, filaments, walls, and voids) can break degeneracies and lead to significantly tighter constraints than the standard matter power spectrum regardless of the scales used to define the cosmic web ($\sim\times80$ tighter constraints on $M_\nu$ in the most optimistic case). While the 3D matter field is not currently directly observable, and weak lensing only measures a 2D projection of the matter field, it may be possible to access this information with tomographic lensing.
    \item While this is true for the total matter field, for the CDM+Baryon field the combined analysis of the power spectra of different cosmic web structures offers far less degeneracy breaking to constrain cosmological parameters -- $M_\nu$ constraints are only $\times3.6$ tighter in the most optimistic case (at this resolution). These modest numbers will apply to galaxy surveys, as galaxies trace the CDM+Baryon field, but nonetheless, a $\times3.6$ increase in constraining power is still of great power to upcoming surveys. This finding is in line with the analysis of \cite{Bayer_2022_fake_nus} which made a simple cross-correlation argument to suggest there is far less information about neutrino mass in the CDM+Baryon field compared to the 3D total matter field.
    \item Adopting a maximum smoothing scale of $\approx 12.5\,{\rm Mpc}/h$ instead of $\approx 2\,{\rm Mpc}/h$ in the cosmic web identification algorithm only degrades the constraints by a factor of $\sim 2-4$, suggesting there is much information in the Hessian of the density field.
\end{itemize}
Looking forward, there are several remaining directions to improve and validate this work. The simplest way to validate this analysis is with higher resolution simulations. Currently there does not exist a suite of simulations that vary all the cosmological parameters explored in this analysis and push to small enough scales where the smoothing scales used to define structures are guaranteed to be converged.

Additionally, we note the impact of redshift-space distortions (RSD) will have a non-trivial impact on the cosmic web distribution which could affect our environmental classification. It was found in a similar analysis \cite{cweb_quijote_Bonnaire_22_2} that multipoles of the power spectrum with cosmic web environment classification on the RSD field helps improve constraints over the real-space power spectrum. In their Table 1 they provide a confusion matrix which shows that the cosmic web structure classification is fairly robust (\~80\%) for structures other than nodes. For these reasons, we save exploration of the impacts of RSD on our analysis for future work.

Furthermore, we note that baryonic feedback is known to have a degenerate small-scale clustering signal with massive neutrinos. We expect baryonic feedback to complicate this analysis and add additional degeneracy which will reduce our constraining power on $M_\nu$ and other cosmological parameters. Currently the CAMELS suite \cite{CAMELS_presentation} is a good potential option for exploring the impacts of baryonic feedback on the cosmic web but the current size of the simulated volume may not be large enough for a robust cosmological analysis. An avenue of future interest with the CAMELS suite would be to explore how the cosmic web could be used to learn information about various hydrodynamical feedback parameters used within the simulations.

Lastly, some additional avenues for future work include exploring cross-correlations between the cosmic web and other observables, such as the CMB, using simulations of the correlated sky such as \cite{Bayer:2024egi}, combining beyond-2pt summary statistics \cite{Beyond-2pt:2024mqz} to improve our cosmological constraints in the future, and using the cosmic web to distinguish different dark matter models \cite{Dome_2023_cweb_dark_matter}.

Alongside our analysis, we have created a new \texttt{python} package \texttt{pycosmommf}\cite{Sunseri_2025_pycosmommf} \footnote{\url{https://github.com/James11222/pycosmommf}} containing a modified version of the \textsc{nexus+}  algorithm. This package has been ported to \texttt{python} from it's original language \texttt{julia} described in our previous work \cite{baryon_Sunseri_23_CosmoMMF}. 

\begin{acknowledgments}
We thank Francisco Villaescusa-Navarro for his extensive assistance with the Quijote Simulations. We thank KG Lee, Zack Li, Daniela Galárraga-Espinosa, and Hideki Tanimura for useful discussions. We are grateful for the stimulating discussions with participants of the Kavli IPMU ``CD3 Hack Fridays'' in Summer 2023, during which the majority of this work was carried out. JS acknowledges support from the Fannie \& John Hertz Foundation. This work was supported by JSPS KAKENHI Grants 23K13095 and 23H00107 (to JL). This work used the Anvil cluster at Purdue University through allocation AST140041 from the Advanced Cyberinfrastructure Coordination Ecosystem: Services \& Support (ACCESS) program, which is supported by National Science Foundation grants \#2138259, \#2138286, \#2138307, \#2137603, and \#2138296. Additionally we used the Princeton Research Computing resources at Princeton University which is a consortium of groups led by the Princeton Institute for Computational Science and Engineering (PICSciE) and Research Computing.
\end{acknowledgments}

\newpage 
\bibliographystyle{physrev}
\bibliography{main}

\newpage

\appendix
\subsection{Numerical Systematics}
\label{subsec:A1}
\begin{figure*}
    \centering
    \includegraphics[width=0.7\linewidth]{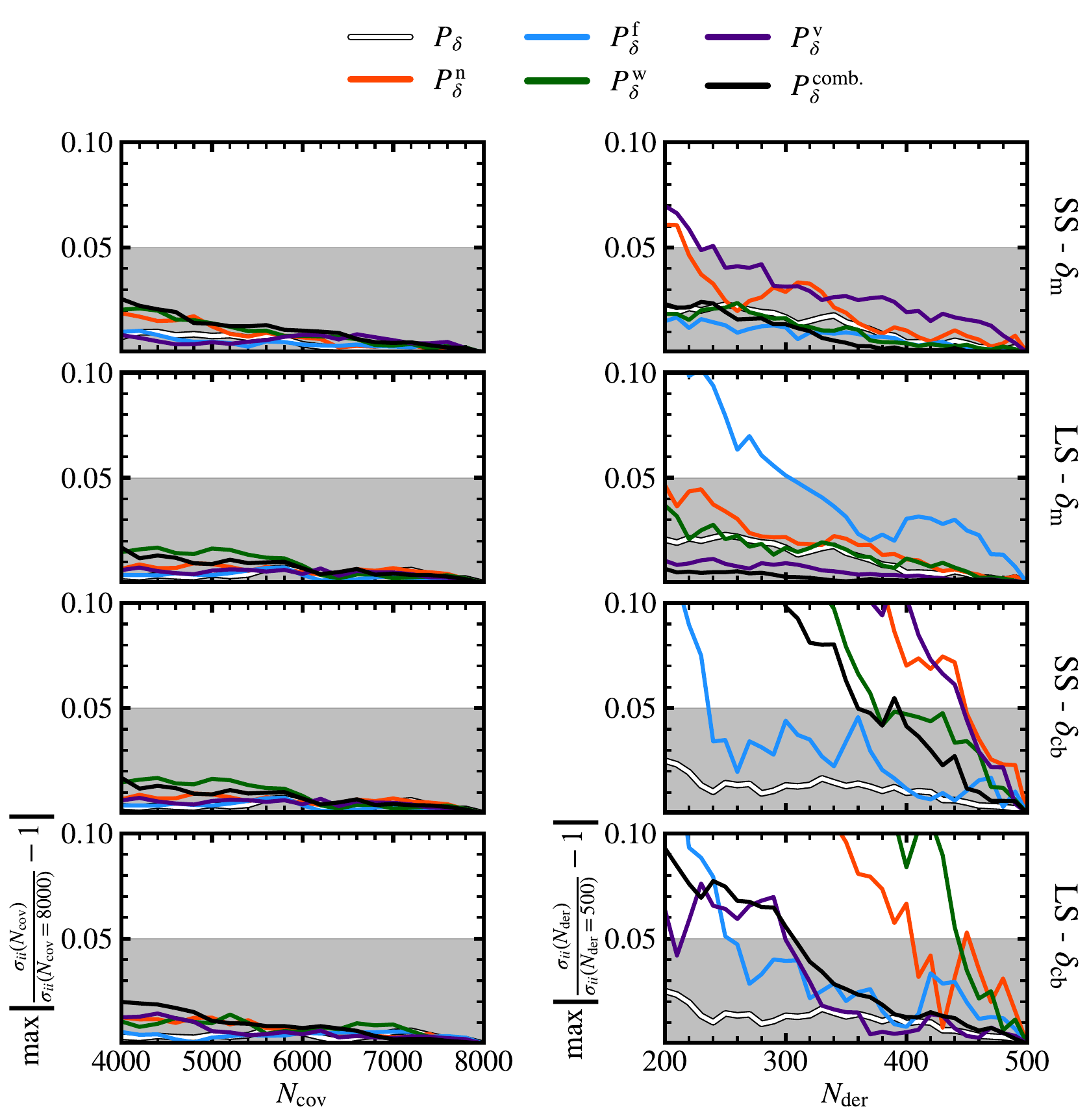}
    \caption{Convergence of errors from all analyses. The left column contains the maximum relative error as a function of $N_{\rm cov}$ compared to an analysis with $N_{\rm der} = 500$ and $N_{\rm cov} = 8000$, the right column is the same except as a function of $N_{\rm der}$ instead. We show convergence for the full power spectrum (white), nodes (red), filaments (blue), walls (green), voids (purple), and all structures combined (black). Each row denotes a different choice of smoothing scales used in the identification scheme --- Small Scales (SS) or Large Scales (LS) and $\delta_{\rm m}$ or $\delta_{\rm cb}$ (probed with galaxy surveys) --- ranging from least conservative to most conservative going from top to bottom.}
    \label{fig:convergence}
\end{figure*}

\begin{figure*}
    \centering
    \includegraphics[width=0.7\linewidth]{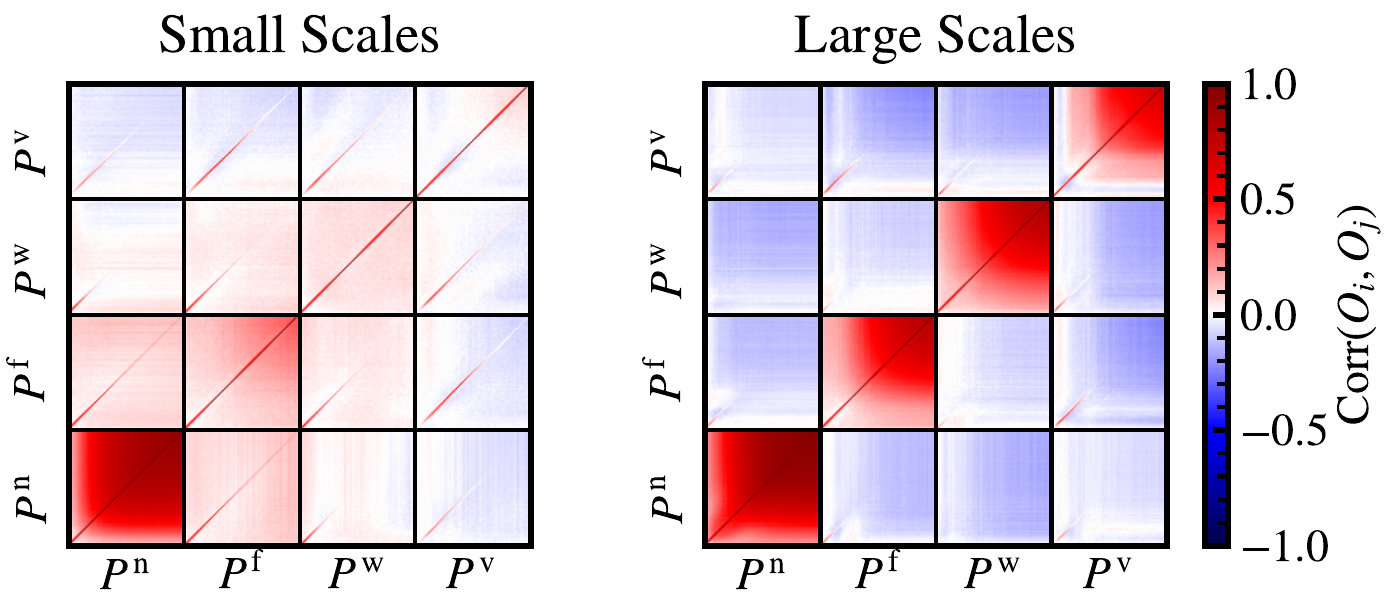}
    \caption{Correlation matrices for the power spectra of nodes, filaments, walls, and voids, each containing 79 scale bins with $k_{\rm max} = 0.5 \; h^{-1} \rm Mpc$. Using small smoothing scales reduces the correlation between structures and radial bins.}
    \label{fig:corr_matrices}
\end{figure*}

In this section we describe the robustness of our results to numerical systematics. We verify that our results are converged by computing the marginalized errors of all parameters as a function of either the number of derivative simulations $N_{\rm der}$ or the number of covariance simulations $N_{\rm cov}$ used. We then compute the maximum fractional difference between these marginalized errors compared to our "converged" analysis with $N_{\rm der} = 500$ and $N_{\rm cov} = 8000$. These can both be written as 
\begin{equation}
    \mathrm{max} \left| \frac{\sigma_{ii}(N_{\rm cov})}{\sigma_{ii}(N_{\rm cov} = 8000)} - 1 \right| \; ,
\end{equation}
and
\begin{equation}
    \mathrm{max} \left| \frac{\sigma_{ii}(N_{\rm der})}{\sigma_{ii}(N_{\rm der} = 500)} - 1 \right| \; .
\end{equation}
The results of this convergence analysis can be found in Figure \ref{fig:convergence}. In all cases the power spectrum of the whole field as a summary statistic is well converged staying well below $5\%$ fractional difference from our "converged" analysis. Our covariance matrices appear well converged for all structures as can be seen in the left hand side of Figure \ref{fig:convergence}. Additionally, we show the correlation matrices $\mathrm{Corr}(O_i, O_j) \equiv C_{i j}/\sqrt{C_{ii} C_{jj}}$ for small and large scale analyses in Figure \ref{fig:corr_matrices}.

We find that our analysis is clearly converged when using the matter field $\delta_{\rm m}$ but this is not as clear for our analyses on the $\delta_{\rm cb}$ field. We do see that there is still some significant fluctuation in marginalized error for individual structures, but we we are confident that our cross-correlation of clustering across all cosmic structures is converged in all cases. Ultimately, the use of additional derivative simulations would help achieve a greater convergence for individual structures but we do not have access to additional derivative simulations as they do not exist. 

\subsection{Contours for all Parameters}
\label{subsec:A2}
\begin{figure*}
    \centering
    \includegraphics[width=0.5\linewidth]{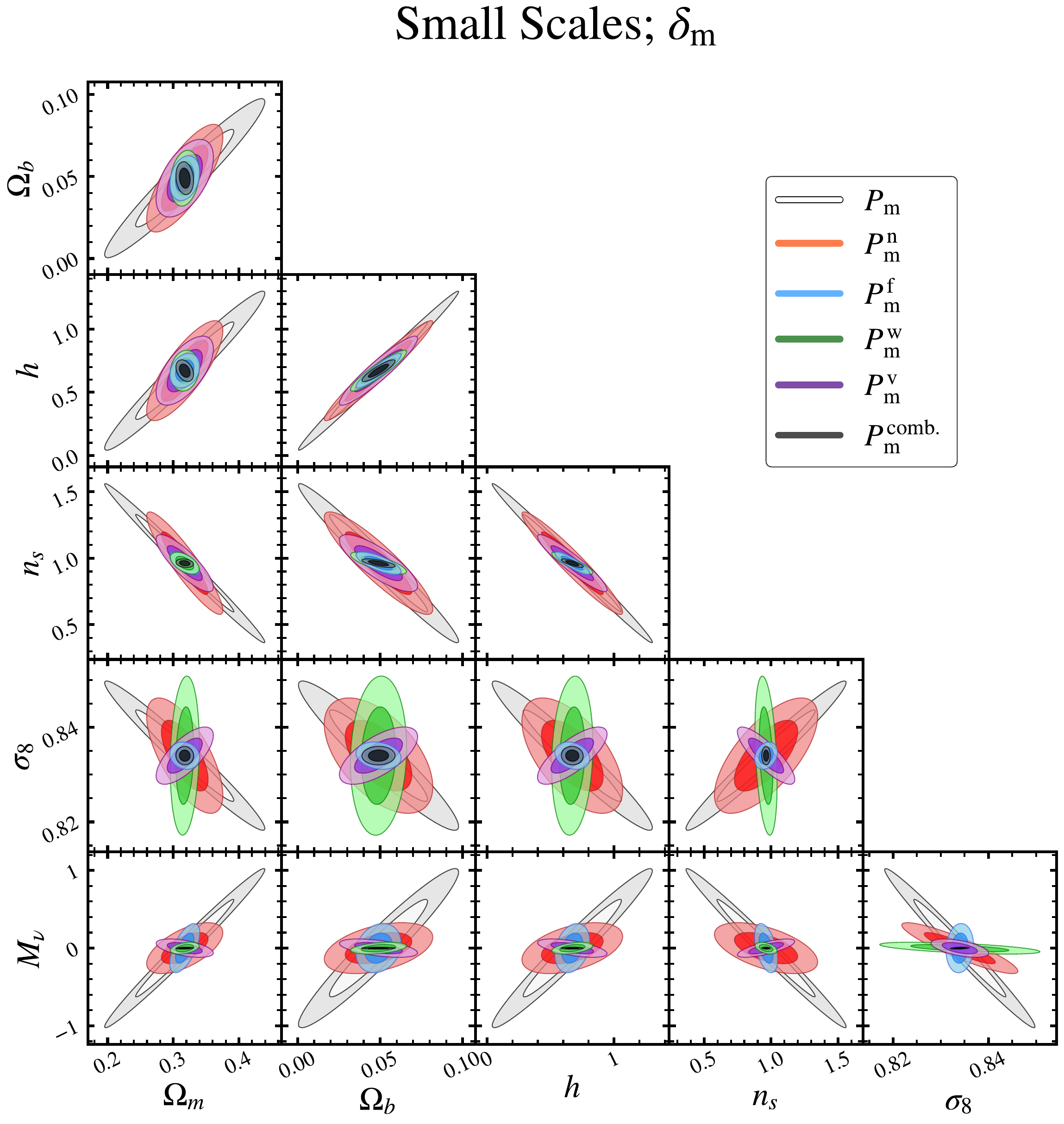}
    \caption{Full forecast of cosmological parameters with the cosmic web. All panels show the $68\%$ (darker shades) and $95\%$ (lighter shades) confidence contours from the standard full power spectrum (white), each structure type --- nodes (red), filaments (blue), walls (green), and voids (purple) --- and the cross correlation of all cosmic structures (black). This figure shows the case of using small smoothing scales on the matter field.}
    \label{fig:Fisher_SS_M}
\end{figure*}

\begin{figure*}
    \centering
    \includegraphics[width=0.5\linewidth]{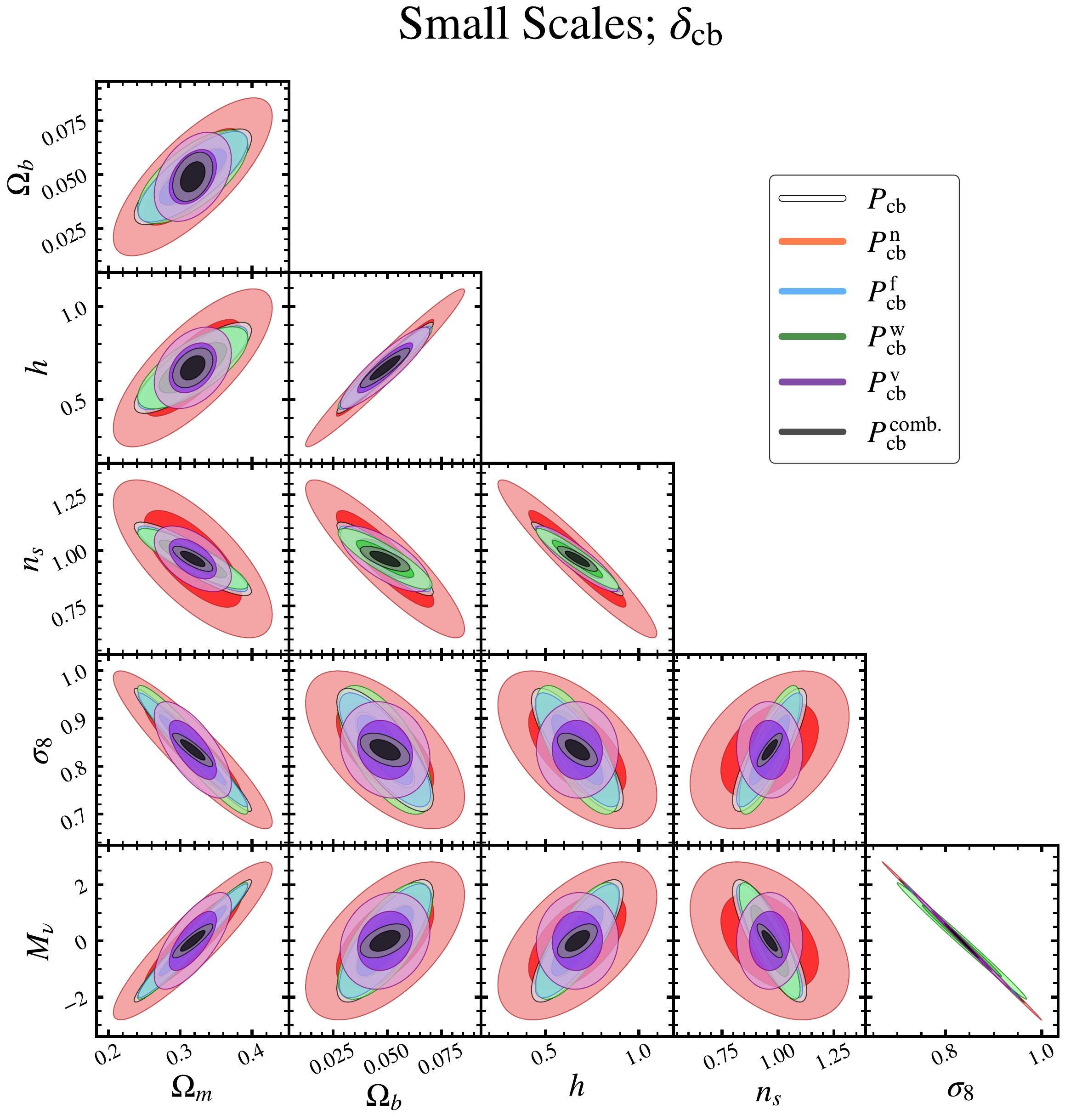}
    \caption{Same as Fig. \ref{fig:Fisher_SS_M} but the case of using small smoothing scales on the $\delta_{\rm cb}$ field.}
    \label{fig:Fisher_SS_CDM}
\end{figure*}

\begin{figure*}
    \centering
    \includegraphics[width=0.5\linewidth]{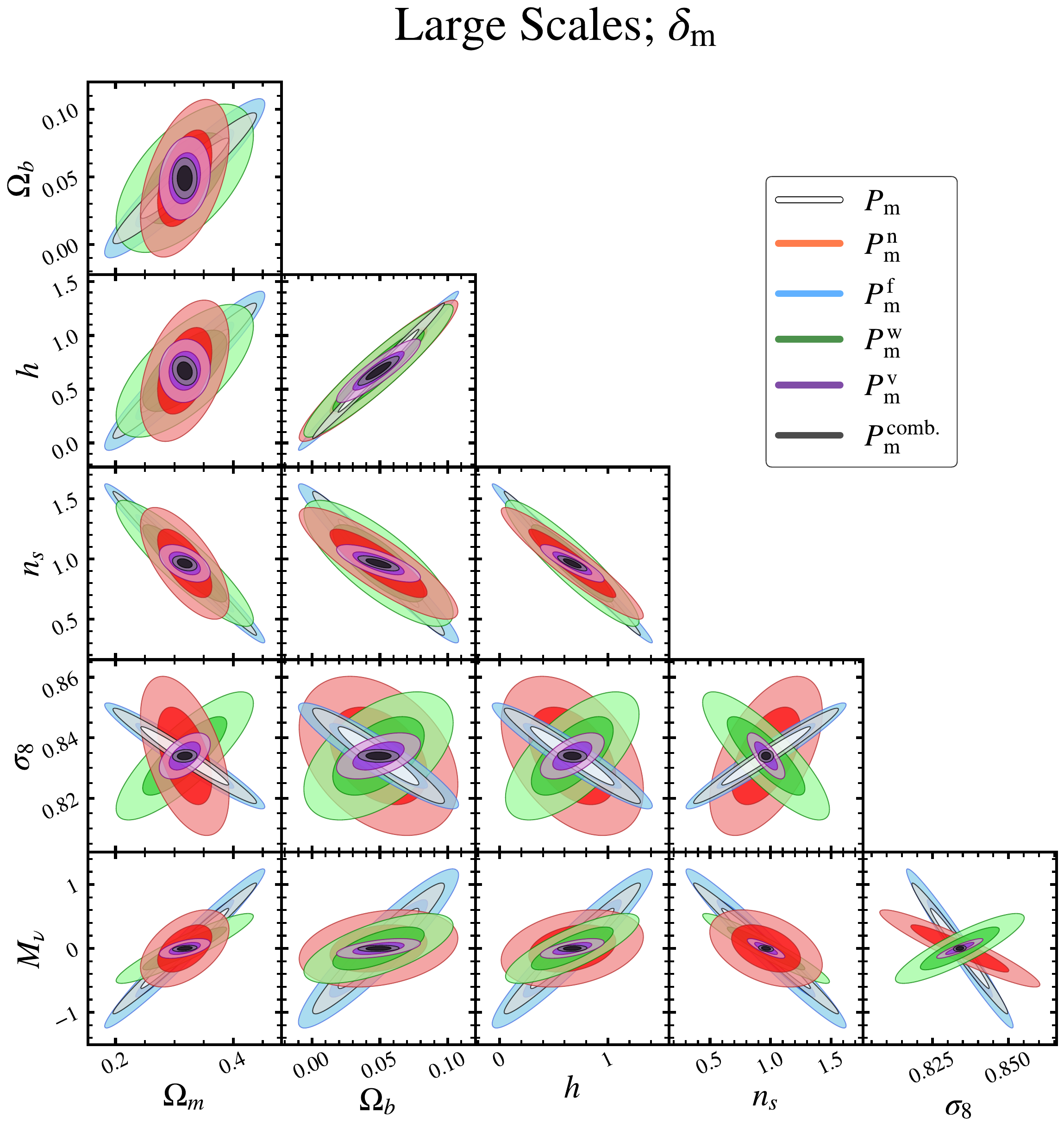}
    \caption{Same as Fig. \ref{fig:Fisher_LS_M} but the case of using large smoothing scales on the matter field.}
    \label{fig:Fisher_LS_M}
\end{figure*}

\begin{figure*}
    \centering
    \includegraphics[width=0.5\linewidth]{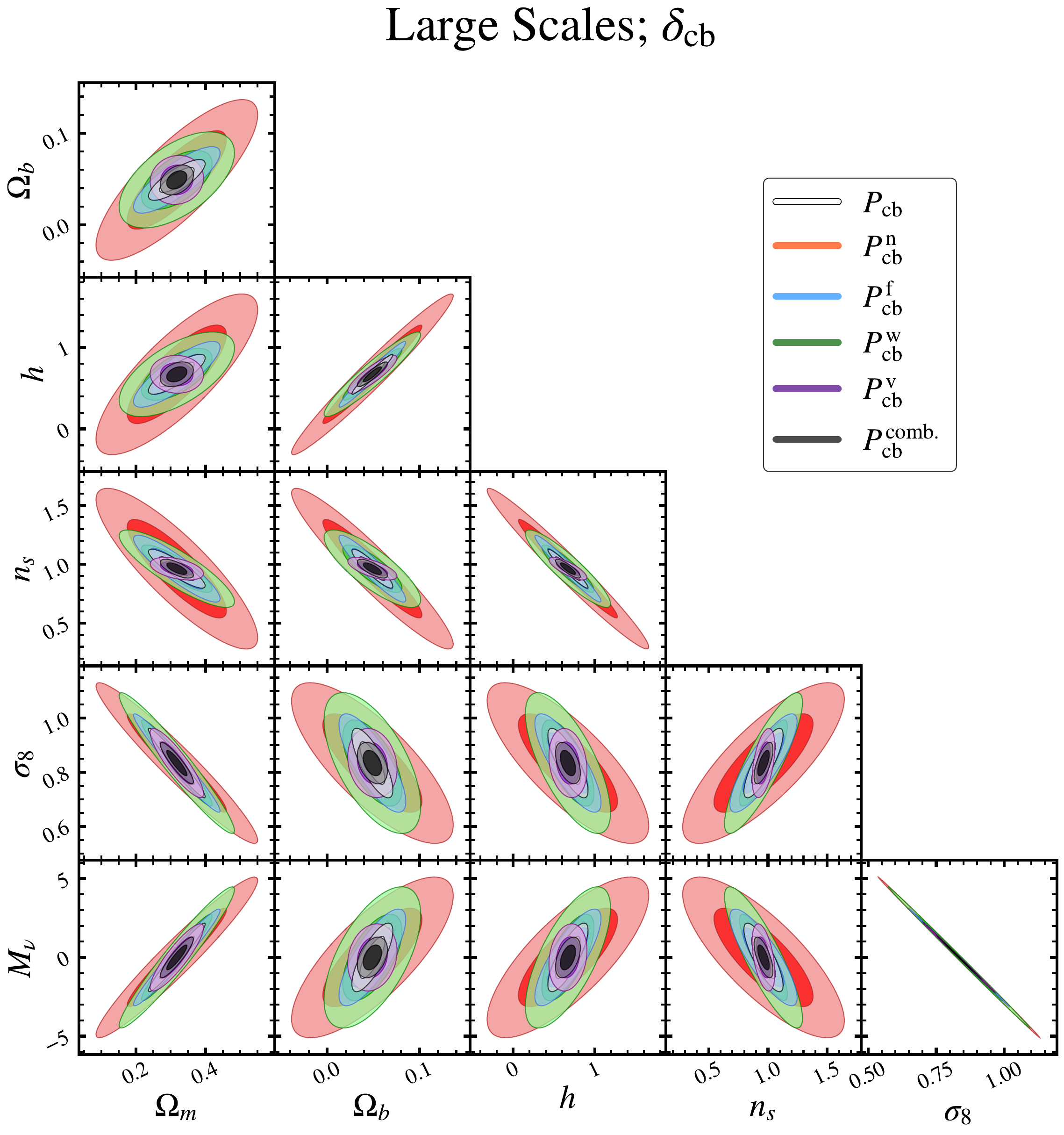}
    \caption{Same as Fig. \ref{fig:Fisher_SS_M} but the case of using large smoothing scales on the $\delta_{\rm cb}$ field.}
    \label{fig:Fisher_LS_CDM}
\end{figure*}

Figures \ref{fig:Fisher_SS_M}, \ref{fig:Fisher_LS_M}, \ref{fig:Fisher_SS_CDM}, and \ref{fig:Fisher_LS_CDM} show contours for the full cosmological parameter space for the four cosmic web analyses described in Section \ref{sec:results}. While the results described in Section \ref{sec:results} focused on the neutrino mass, the general trends are similar for all parameters.

\newpage

\end{document}